\documentclass[
 reprint,
 amsmath,amssymb,
 aps,
 nofootinbib,
 prl,
 superscriptaddress,
 floatfix,
]{revtex4-1}
\usepackage{graphicx}
\usepackage{dcolumn}
\usepackage{bm}
\usepackage{amsmath}
\usepackage{float}
\usepackage{lipsum}
\usepackage{xcolor}
\usepackage{textcomp}
\usepackage{latexsym}
\usepackage[colorlinks=true,citecolor=red,urlcolor=red]{hyperref}
\usepackage{mathtools}
\usepackage{url}
\usepackage{comment}
\usepackage[utf8]{inputenc}
\usepackage[normalem]{ulem}
\usepackage{xspace}
\usepackage{acronym}

\newcommand{\beq}{\begin{equation}}
\newcommand{\eeq}{\end{equation}}
\newcommand{\Msolar}{{\rm M}_{\odot}}
\newcommand{\Mmax}{M_{\rm max}}
\newcommand{\SNR}{\rm SNR}

\newcommand{\LIGOlabMIT}{\affiliation{LIGO Laboratory, Massachusetts Institute of Technology, 185 Albany St, Cambridge, MA 02139, USA}}
\newcommand{\MKI}{\affiliation{Department of Physics and Kavli Institute for Astrophysics and Space Research, Massachusetts Institute of Technology, 77 Massachusetts Ave, Cambridge, MA 02139, USA}}
\newcommand{\CCA}{\affiliation{Center for Computational Astrophysics, Flatiron Institute, 162 5th Ave, New York, NY 10010, USA}}
\newcommand{\GSI}{\affiliation{GSI Helmholtzzentrum f\"ur Schwerionenforschung, Planckstra{\ss}e 1, 64291 Darmstadt, Germany}}
\newcommand{\GeorgiaTech}{\affiliation{Center for Relativistic Astrophysics and School of Physics, Georgia Institute of Technology, Atlanta, GA 30332}}

\begin{document}

\title{Inference of the neutron star equation of state from cosmological distances}

\author{Carl-Johan Haster} \LIGOlabMIT \MKI
\author{Katerina Chatziioannou} \CCA
\author{Andreas Bauswein} \GSI
\author{James Alexander Clark} \GeorgiaTech

\begin{abstract}
Finite-size effects on the gravitational wave signal from a neutron star merger typically manifest at high frequencies where detector sensitivity decreases.
Proposed sensitivity improvements can give us access both to stronger signals and to a myriad of weak signals from cosmological distances. 
The latter will outnumber the former and the relevant part of signal will be redshifted towards the detector most sensitive band. 
We study the redshift dependence of information about neutron star matter and find that single-scale properties, such as the star radius or the post-merger frequency, are better measured from the distant weak sources from $z\sim 1$.
\end{abstract}

\maketitle


\acrodef{SNR}[SNR]{signal-to-noise ratio}
\acrodef{BH}[BH]{black hole}
\acrodef{BBH}[BBH]{binary black hole}
\acrodef{BNS}[BNS]{binary neutron star}
\acrodef{NS}[NS]{neutron star}
\acrodef{GW}[GW]{gravitational wave}
\acrodef{EoS}[EoS]{equation of state}
\acrodef{CE}[CE]{Cosmic Explorer}
\acrodef{ET}[ET]{Einstein Telescope}
\acrodef{aLIGO}[aLIGO]{Advanced LIGO}


\section{Introduction}
\label{sec:intro}

The \ac{GW} signal emitted during the coalescence of two \acp{NS} carries the imprint of the \ac{EoS} of cold, supranuclear matter~\cite{Faber2012,Baiotti:2016qnr}. 
Extracting \ac{EoS} attributes is a key science goal not only of \ac{GW} astronomy, but also of diverse probes such as nuclear experiment and theory, and radio and X-ray observations of pulsars~\cite{Lattimer_2005,Antoniadis:2013pzd,Miller:2016pom,Ozel:2016oaf,Cromartie:2019kug}. 
To date, \acp{GW} from two \ac{BNS} coalescences have been detected~\cite{TheLIGOScientific:2017qsa,Abbott:2018wiz,Abbott:2018exr,Abbott:2020uma} and combined with other constraints to study the \ac{EoS}, e.g.~\cite{Margalit:2017dij,Bauswein:2017vtn,Abbott:2018exr,Coughlin:2018fis,Raaijmakers:2019dks,Landry:2020vaw}. 
Projections from simulated observations suggest \acp{GW} will dominate the astrophysical \ac{EoS} constraints in the next $\sim5$ years~\cite{Landry:2020vaw}.

Ground-based \ac{GW} detectors observe a \ac{BNS} coalescence signal for multiple minutes, starting at $\sim10-20\mathrm{Hz}$ and sweeping up in frequency towards merger. 
The lower frequency cutoff is predominantly determined by the detector sensitivity.  
The final, high frequency part of the signal nominally falls within the detector bandwidth, but depending on the system mass and \ac{EoS} the decreased detector sensitivity might make its accurate extraction challenging in the near future. 
The signal up to a few hundred Hz is not expected to inform \ac{NS} \ac{EoS} constraints as the inspiraling bodies are sufficiently separated such that matter effects are subdominant (excluding potential resonant effects~\cite{Pratten:2019sed,Ma:2020rak}, instabilities~\cite{Weinberg:2015pxa, Weinberg:2018icl}, etc.); they scale as $(R/r)^5$, where $R$ is the \ac{NS} radius and $r$ the binary separation~\cite{Flanagan:2007ix}.

Once the \acp{NS} are sufficiently close and the signal reaches a few hundred Hz, mutual tidal interactions induce quadrupole moments on each star.
The additional quadrupole moment (besides the orbital one) affects both the binding energy and the rate of energy extraction, effectively speeding up the system evolution towards merger. 
This speed-up depends on the \acp{NS} size (less compact \acp{NS} are generally more deformable) and can be measured from the phase evolution of the signal, constraining the \acp{NS} deformability. 
Depending on the \ac{NS} mass and radius, the part of the signal influenced by tidal interactions corresponds to frequencies $\gtrsim 400\mathrm{Hz}$ in the source frame of the \ac{BNS}. 
After the inevitable collision, and unless it promptly forms a BH, the merger remnant emits a signal dominated by a single frequency component, usually at $1500-4000\mathrm{Hz}$~\cite{Xing:1994ak,Shibata:2005ss,shibata:06bns,Oechslin:2007gn,2011PhRvD..83l4008H,2012PhRvL.108a1101B,Bauswein:2012ya,hotokezaka:13,2014PhRvL.113i1104T,2015arXiv150401764B,bauswein:15,Foucart2016,Lehner:2016lxy,2016CQGra..33x4004E,Dietrich2017}.
This mode has been linked to f-modes of the remnant star~\cite{Stergioulas:2011gd,Bauswein:2015vxa}; it depends on the remnant's density, so it can offer complementary information about the \ac{EoS} stiffness. 
For more massive \acp{NS} or for stiffer \acp{EoS} the dominant frequency is lower and the signal is easier to extract from the noise~\cite{Bauswein:2012ya}.

The two observed \ac{BNS} signals already constrain tidal interactions between the coalescing \acp{NS}~\cite{Abbott:2018exr,Abbott:2018wiz,De:2018uhw,Dai:2018dca}, but the post-merger signal was lost in the detector noise~\cite{Abbott:2018wiz,Abbott_2017}. 
Planned improvements of \ac{GW} detectors~\cite{Aasi:2013wya,Lynch:2014ffa,Tse:2019wcy,McCuller:2020yhw} will not only allow for better tidal measurements, but might also reveal the elusive post-merger signal~\cite{Yang:2017xlf,Bose:2017jvk,Chatziioannou:2017ixj,Torres-Rivas:2018svp,Martynov:2019gvu}. 
Further ahead, next generation detectors~\cite{Evans:2016mbw, Reitze:2019dyk,Reitze:2019iox,Punturo:2010zz,Maggiore:2019uih} predict a $\mathcal{O}(10)$ sensitivity increase, potentially detecting and individually resolving a significant fraction of all \ac{BNS} coalescences even at cosmological distances~\cite{Sachdev:2020bkk}, occurring every ${\mathcal{O}}(10\mathrm{s})$~\cite{Abbott:2017xzg}.

\begin{figure}[h]
\includegraphics[width=\columnwidth,clip=true]{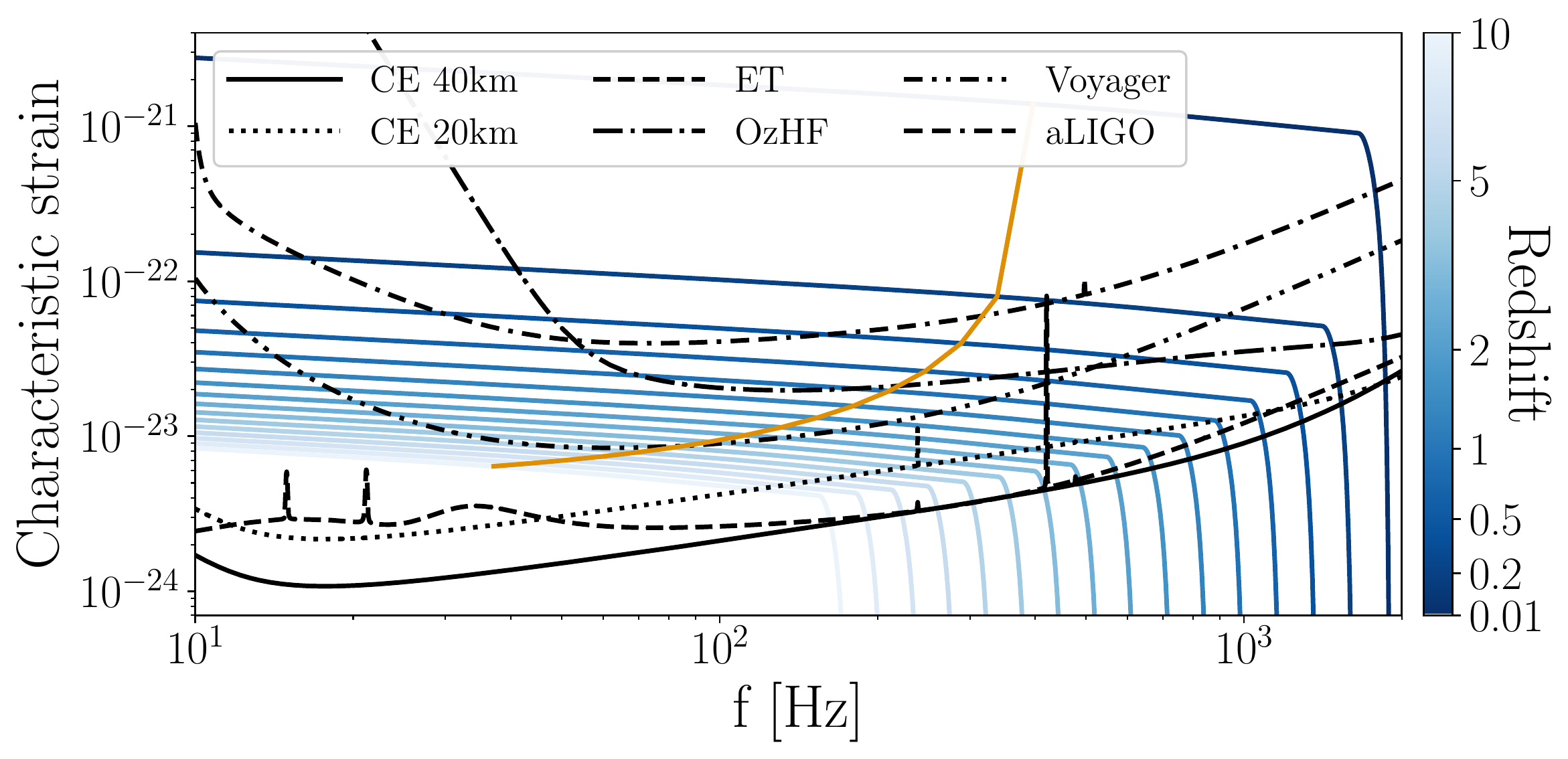}
\includegraphics[width=\columnwidth,clip=true]{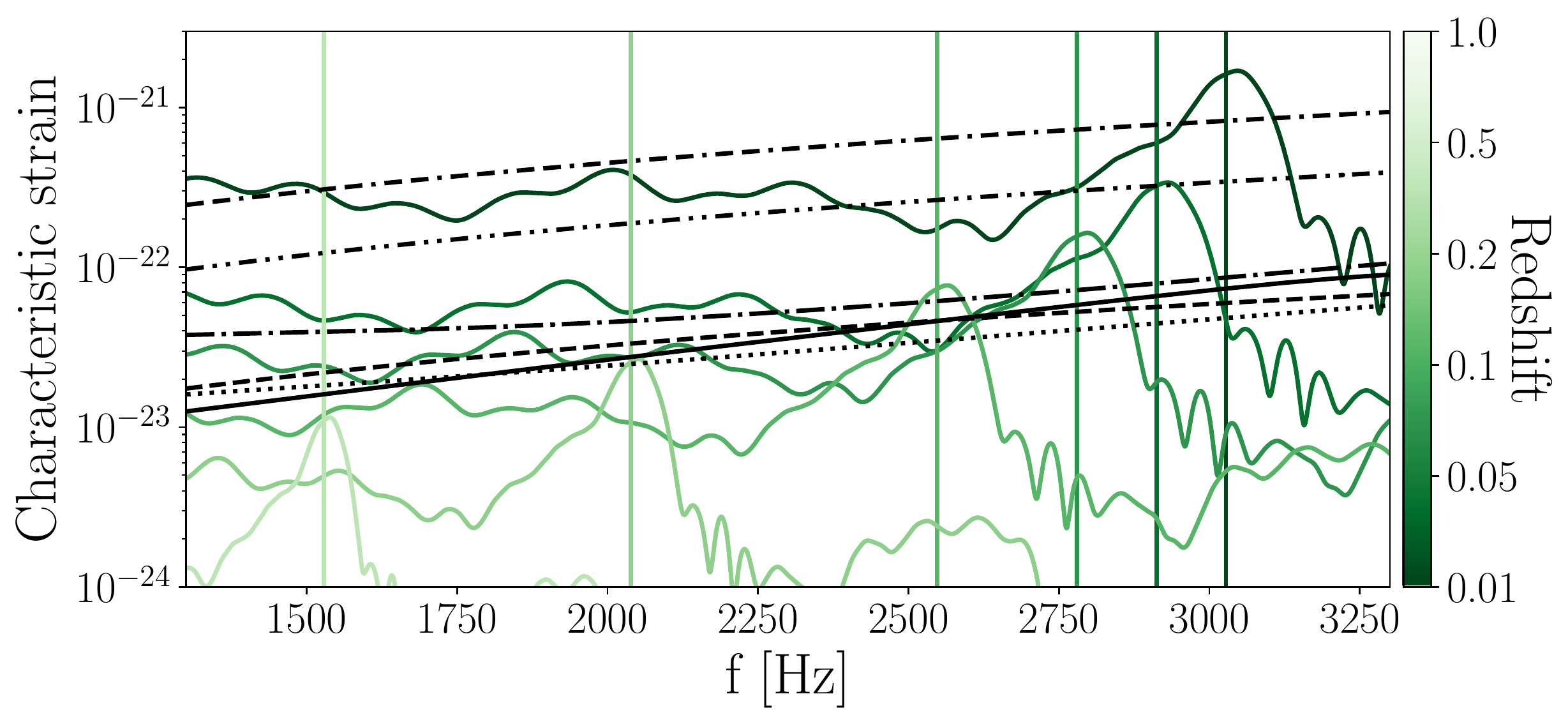}
\caption{Charachteristic strain~\cite{Moore:2014lga} of a source-frame $1.362 + 1.362 \Msolar$ \ac{BNS} signal at different redshifts (shaded blue or green) and planned or proposed detectors (black lines).
\textit{Top panel}: Late-inspiral signal assuming the H4 \ac{EoS}~\cite{Lackey:2005tk}. 
The orange line represents $400\mathrm{Hz}$ in the source frame; the tidally-corrected signal is always observed with $f>10$Hz.
\textit{Bottom panel}: Post-merger signal assuming EoS7 from~\cite{Torres-Rivas:2018svp}. 
Vertical lines mark the dominant post-merger frequency.
}
\label{fig:PSDplot}
\end{figure}

The cosmological reach of next generation detectors results in observations with non-negligible redshifting compared to the emitted signal. 
Cosmological redshifting is familiar in \ac{GW} astronomy as even current-sensitivity \ac{BBH} signals are appreciably redshifted~\cite{LIGOScientific:2018mvr}.
This can be a blessing and a curse for \acp{BBH} as redshifting simultaneously amplifies the signal (since masses appear larger) and shifts it to lower frequencies, potentially outside the detectors' sensitive band. 
Observed \ac{BNS} signals will also appear more massive and lower in frequency, resulting in some early inspiral signal being lost to low-frequency noise. 
However, for the typical \ac{BNS} reach of next generation detectors, $z\sim 2-3$, this early signal is generally not informative about finite-size effects. 
The late inspiral and post-merger signals are typically emitted above $400\mathrm{Hz}$, so even a serendipitous detection at $z=10$ would redshift the signal to $\gtrsim 35\mathrm{Hz}$, safely above the noise dominated frequencies. 
Figure~\ref{fig:PSDplot} shows the observed pre-merger (top) and post-merger (bottom) \ac{BNS} signal at different redshifts compared to various detector noise spectra. 
Both signals move to lower frequencies with increasing redshift, but always above $10$Hz.

Besides redshifting, distant sources are also more numerous due to volume effects. 
The number of binaries per time, per redshift slice in the observer frame is
\begin{equation}
R(z)=\frac{dN}{dT dz}=\frac{R_s(z)}{1+z}\frac{dV_c}{dz},
\label{eq-zdistribution}
\end{equation}
where $dV_c/dz$ is the differential comoving volume, $R_s(z)$ is the rate per comoving volume in the source frame, and the $1+z$ term converts from the source frame to the observer frame. 
We assume $R_s(z)$ has the same shape as the star formation rate~\cite{Madau:2014bja}, which peaks at $z\sim 2$ and corresponds to an effective zero delay time between binary formation and merger. 
A nonzero delay time would smoothen the redshift-dependent rate, making it less peaked than the star formation rate.
We assume a local \ac{BNS} merger rate as $R_s(z=0)=500$Gpc${}^{-3}$yr$^{-1}$, consistent with current BNS observations~\cite{Abbott:2020uma}.

We study whether the fewer, loud, close-by signals or the numerous, weak, distant, redshifted signals offer the most information about the \ac{NS} \ac{EoS}. 
Expectedly, the answer depends on the detector cosmological reach. 
Next-generation detectors can extract a single \ac{EoS} scale, such as the \ac{NS} radius at a fiducial mass or the frequency of the post-merger oscillations, 
better with the multitude of far-away sources, thanks to appreciable nonlocal information. 
Conversely, detailed features of the \ac{EoS} such as the functional dependence of tidal properties and radius on the mass require the simultaneous extraction of multiple parameters and constraints are dominated by the few loud sources from the local Universe.

\section{Pre-merger signal}
\label{sec:premerger}

During the late inspiral phase of a \ac{BNS} coalescence, matter effects are encoded in the parameter $\Lambda$ quantifying the tidal deformation of a \ac{NS} under an external field~\cite{Flanagan:2007ix,Favata:2013rwa,Wade:2014vqa}. 
We estimate the expected information gained from measuring the tidally-corrected signal and $\Lambda$ through the Fisher information matrix $F$~\cite{Vallisneri:2007ev}. 
The accumulated information can be approximated by ${\rm det} F$, the determinant of $F$ which scales with the observed \ac{SNR} as ${\rm det} F \sim \SNR^{2D}$, where $D$ is the number of relevant parameters~\cite{Becsy:2019dim}.
For $D=1$ this reduces to the well known SNR$^2$ dependence on the measurement covariance and the expectation that SNR adds in quadrature rather than linearly. 

$\Lambda$ depends on the \ac{NS} mass $m$, so to accurately charachterize a generic $\Lambda(m)$ curve, $D$ is in principle large.
In practice, generic \acp{EoS} can be approximated with a few parameters only. 
For hadronic \acp{EoS}, only a single parameter is expected to be measurable with second generation detectors, for example $\Lambda(m=1.4\Msolar)$~\cite{DelPozzo:2013ala}. 
Such a measurement would only provide a single scale of the \ac{EoS}, such as the radius at a specific \ac{NS} mass. 
\ac{EoS}-insensitive relations also suggest that hadronic \acp{EoS} can be suitably described by a single parameter~\cite{Yagi:2015pkc, Chatziioannou:2018vzf, Raithel:2018ncd, De:2018uhw, Kumar:2019xgp}. 
Instead, exploring features of the \ac{EoS}, such as the mass dependence of the tidal parameter or the presence of a phase transition~\cite{Oertel:2016bki,MontanaTolos2019, Essick:2019ldf, Chen:2019rja, Chatziioannou:2019yko}, relies on \ac{EoS} parametrizations. 
Common models employ $\sim4$ or more parameters~\cite{Read:2008iy,Lackey:2014fwa,Lindblom:2018rfr,Carney:2018sdv,Raithel:2016bux,Tews:2018kmu,Greif:2018njt} with additional parameters for eventual phase transitions~\cite{Alford:2015gna,Han:2018mtj,Greif:2018njt}. 
With this in mind we present results for $D=\{1,2\}$.

\begin{figure}[h]
\includegraphics[width=\columnwidth,clip=true]{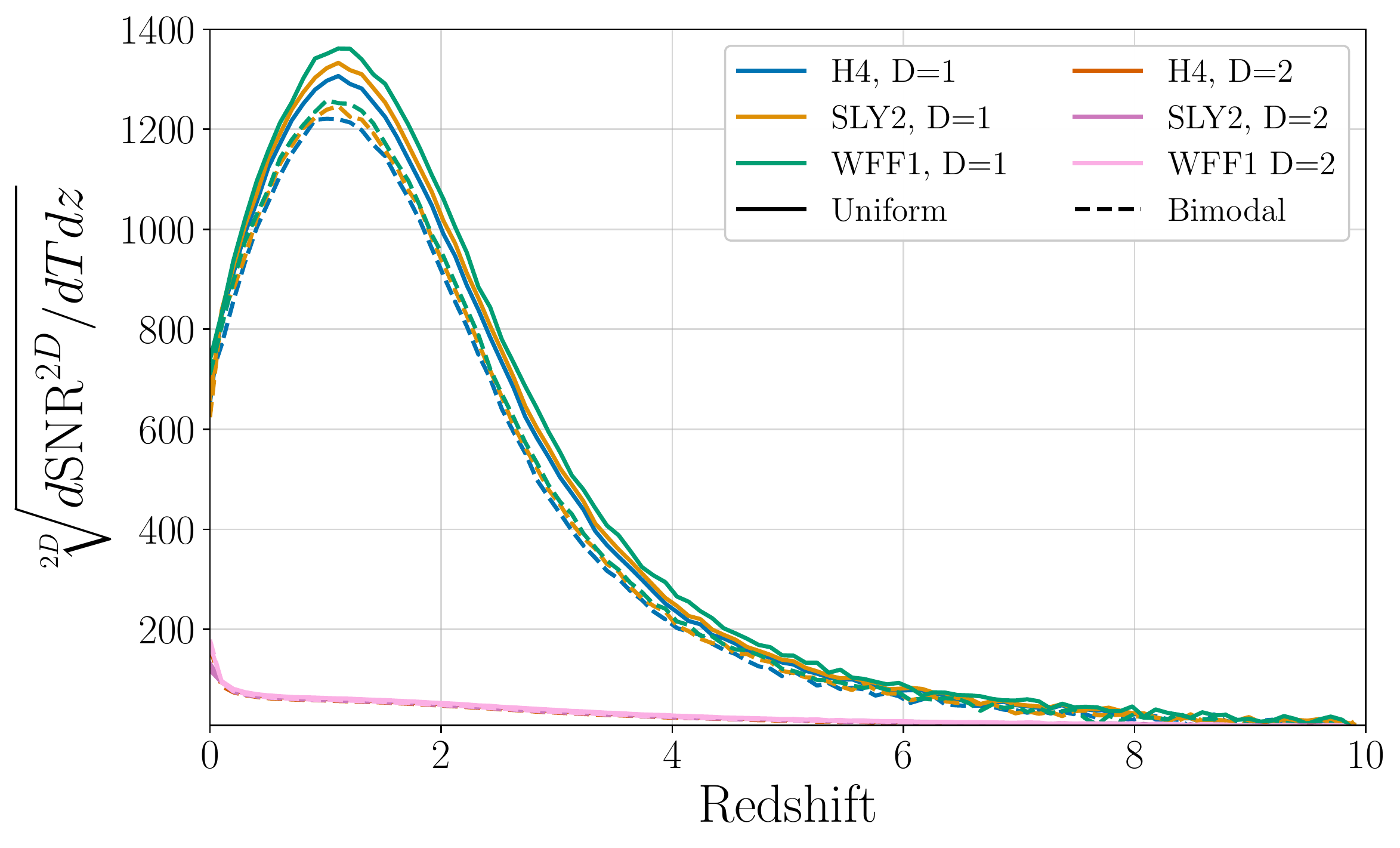}
\caption{
Total pre-merger \ac{SNR} distribution per year per redshift slice, as a function of redshift for two mass distributions, three \acp{EoS}, $D=\{1,2\}$, and the $40\mathrm{km}$ \ac{CE} detector. 
For $D=1$, the availableSNR is dominated by \acp{BNS} at cosmological distances.
For $D=2$, the accumulated information peaks at $z\sim 0$, suggesting a main contribution from local sources.
}
\label{fig:SNRpre}
\end{figure}

We simulate signals from one year's worth of nonspinning \ac{BNS} coalescences with the redshift distribution of Eq.~\ref{eq-zdistribution}. 
We use WFF1~\cite{Wiringa:1988tp}, SLY2~\cite{Gulminelli:2015csa,Danielewicz:2008cm}, and H4~\cite{Lackey:2005tk} as fiducial \acp{EoS} with varying stiffness, but consistent with GW170817~\cite{Abbott:2018exr, LIGOScientific:2019eut}. 
We assume two source frame mass distributions: (i) masses are drawn uniformly in $[1,\Mmax]\Msolar$, where $\Mmax$ is the assumed \ac{EoS} maximum mass, and (ii) the primary mass is drawn from a bimodal distribution~\cite{Alsing:2017bbc} with a mass ratio $q$ distribution $\sim q^3$~\cite{Dominik:2012kk}. 
We distribute systems uniformly across the sky and binary orientations and retain systems with single-detector \ac{SNR} above 8, evaluated from $10\mathrm{Hz}$ with the waveform model \textsc{IMRPhenomPv2\_NRTidal}~\cite{Dietrich:2018uni} and the proposed $40\mathrm{km}$ \ac{CE} instrument~\cite{Evans:2016mbw, Reitze:2019dyk,Reitze:2019iox}(CE2 from~\cite{CosmicExplorer}).
For \ac{EoS} inference we only consider the \ac{SNR} above $400\mathrm{Hz}$ in the source frame (orange line in Fig.~\ref{fig:PSDplot}) restricting to the tidally-corrected signal. 
Figure~\ref{fig:SNRpre} shows the total information per redshift slice per year for $D=\{1,2\}$; at each redshift bin we add SNR$^{2D}$ from detectable binaries, collecting the total information by systems at that distance.

For the inspiral signal in the stationary phase approximation~\cite{Droz:1999qx}, the \ac{SNR} is proportional to $\mathcal{M}^{5/6}/D_L\sim (1+z)^{5/6}/D_L$, where $D_L$ is the luminosity distance and $\mathcal{M} \equiv (m_1 m_2)^{3/5}(m_1 + m_2)^{-1/5}$ is the source-frame chirp mass. 
As $D_L$ increases, the denominator increases linearly; distant signals are weaker.
Simultaneously, the redshifted mass is higher than the source-frame mass resulting in a higher signal amplitude and there are more sources at large distances; the $\SNR^2$ distribution peaks at  $z\sim1$. 
The $\SNR^4$ distribution, on the other hand, is a monotonically decreasing function, as the $\sim 1/D_L^4$ decline cannot be compensated by either redshifting or volume effects. 
For $D=1$, corresponding to measuring a single EoS scale, information is dominated by events at $z\sim1$.
For $D>1$ the distribution peaks at $z=0$ and rare loud events dominate, consistent with~\cite{Lackey:2014fwa}.

Besides the differing shapes of the distributions, the total accumulated \ac{SNR} is also higher for $D=1$ for all \acp{EoS} and mass distributions.
For a set of $N$ observations, the total \ac{SNR} grows as $N^{1/(2D)}$, suggesting a more efficient collection of information for lower-dimensional analyses.
The effect of the assumed mass distribution and \ac{EoS} is smaller, though as expected we find that the uniform mass distribution leads to larger \acp{SNR} as it contains a larger fraction of massive \acp{NS}.
Stiff \acp{EoS} -- such as H4 -- predict large \ac{NS} radii for a given mass and hence an earlier tidal disruption relative to a softer \ac{EoS}, corresponding to a 
small reduction in \ac{SNR}.

\section{Post-merger signal} \label{sec:postmerger}

The post-merger signal is dominated by a single frequency component, see Fig.~\ref{fig:PSDplot}, that depends on the \ac{EoS} stiffness. 
Subdominant structure also exists~\cite{Stergioulas:2011gd, Takami:2014zpa}, but the main subdominant modes are related to the dominant one~\cite{bauswein:15,Clark:2015zxa}. 
We therefore use $D=1$ and follow a similar procedure as above, using only systems detectable by their pre-merger signal. 
For the detected post-merger signal we employ equal-mass simulations with EoS6, EoS7, and EoS4 from~\cite{Torres-Rivas:2018svp}, constructed to be consistent with GW170817~\cite{Abbott:2018exr}. 
Simulations were performed with a relativistic smooth particle hydrodynamics code, with conformal flatness to solve the Einstein equations~\cite{Wilson:1996ty,Oechslin:2001km,Oechslin:2006uk,Bauswein:2010dn}.

We assume the same mass and orientation distribution as above. 
The amplitude of the post-merger signal is generally expected to increase with mass and drop steeply when the total mass $M$ approaches the threshold for prompt BH formation.
Reference~\cite{Tsang:2019esi} presents a phenomenological model for the post-merger signal with the amplitude proportional to $M$ though with some residual dependence on $q$~\cite{Tsang:2019esi,Breschi:2019srl,Bernuzzi:2020txg}. 
Our signals are based on numerical simulations with a certain $M$~\cite{Torres-Rivas:2018svp}; to estimate the observed signal from a binary with a different $M$ and $z$, we scale the amplitude proportionally to the redshifted $M$~\cite{Tsang:2019esi}. We also adjust the dominant
 frequency inversely proportionally to the system's redshifted $M$. 
Finally, systems whose mass exceeds an \ac{EoS}-dependent threshold result in prompt collapse to a BH, emitting negligible signal. 
We use the relations in~\cite{Bauswein:2013jpa,Bauswein:2020aag} to quantify this threshold for each EoS 
and discard systems exceeding it by setting their post-merger \ac{SNR} to zero.

\begin{figure}[h]
\includegraphics[width=\columnwidth,clip=true]{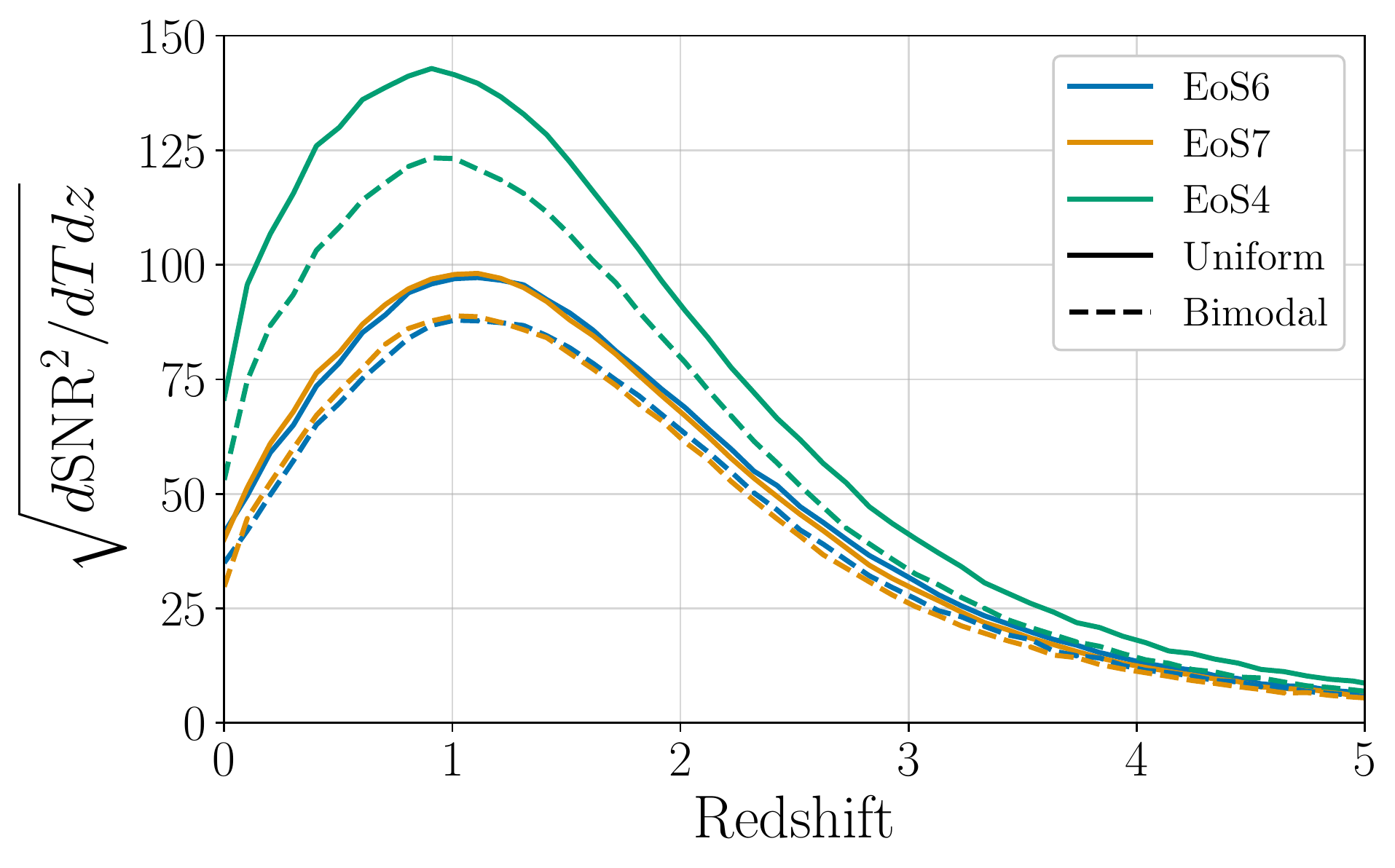}
\caption{
Total post-merger \ac{SNR} distribution per year per redshift slice as a function of redshift for two mass distributions, three \acp{EoS} and the $40\mathrm{km}$ \ac{CE} detector ($D=1$).
The distribution peaks at $z\sim1$, suggesting a significant contribution from nonlocal sources on the total post-merger information.
}
\label{fig:SNRpost}
\end{figure}

Figure~\ref{fig:SNRpost} shows the total information as a function of the redshift for all \acp{EoS} and mass distributions. 
We find that the available post-merger SNR is dominated by nonlocal sources with a distribution peaking at $z\sim1$. 
The uniform mass distribution results in overall larger detected \ac{SNR}, as it results in heavier systems and louder signals. 
Additionally, stiffer \acp{EoS} generally result in a higher total \ac{SNR} as their post-merger frequency is lower. 
However, the threshold mass is also important, as a low threshold to prompt collapse means that the heavier (and hence potentially louder) systems emit no relevant signal. 
As a result of this tradeoff, the stiffest \ac{EoS} in our set, EoS4, indeed results in the highest \ac{SNR}, but there is no clear distinction between EoS6 and EoS7.

Finally, we comment on a phenomenon observed in~\cite{Torres-Rivas:2018svp}. 
Depending on the shape of the post-merger spectrum and the steepness of the detector noise, the lower-frequency subdominant modes might result in higher \ac{SNR} than the dominant one, leading to a ``reversal" in which is observed first. 
Combining multiple sources at different masses and redshifts could help mitigate any potential confusion arising from a misidentification of the dominant mode from a loud source.

\section{Effect of Detector configuration} 

The conclusion that nonlocal binaries contribute significant information about finite-size effects hinges on detectors with cosmological reach for \acp{BNS} that can detect signals out to redshift $z\gtrsim 1$. 
To explore this effect, we repeat the above analyses with five additional detector configurations, see Fig.~\ref{fig:PSDplot}.
Besides the $40\mathrm{km}$ \ac{CE} detector, we also study its $20\mathrm{km}$ version~\cite{Evans:2016mbw, CE_20} with nominally greater sensitivity at frequencies above 2kHz~\cite{CE_mirror_spec} compared to the $40\mathrm{km}$ version.
We also include the \ac{ET}~\cite{Punturo:2010zz,Maggiore:2019uih,ASD_curves18} (in its ET-D configuration), a LIGO Voyager detector~\cite{Adhikari:2020gft, ASD_curves18} and \ac{aLIGO} at its final design sensitivity~\cite{TheLIGOScientific:2014jea,aLIGO_design_updated}.
Finally, we include OzHF~\cite{Bailes:2019oma, OzHF}, a proposed detector designed for dedicated high-frequency \ac{GW} observations.

\begin{figure}[h]
\includegraphics[width=\columnwidth,clip=true]{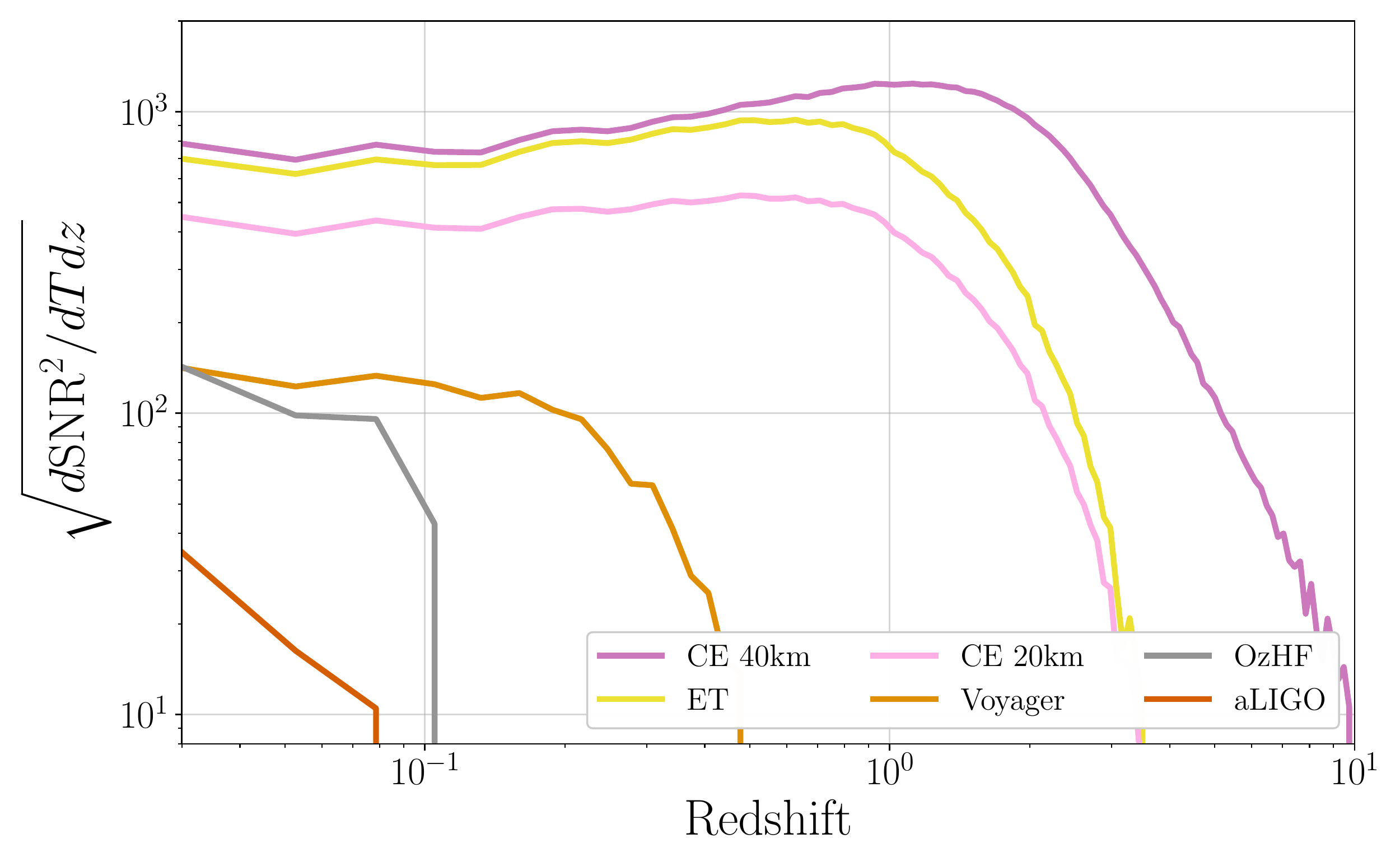}
\caption{
Total pre-merger \ac{SNR} distribution per year per redshift slice, as a function of redshift for the bimodal mass distribution, the H4 \ac{EoS} and for different planned or proposed detectors. 
We show results for $D=1$.
}
\label{fig:SNRpre_manyIFO}
\end{figure}

The total pre-merger information with each detector configuration is shown in Fig.~\ref{fig:SNRpre_manyIFO}, for a single-parameter analysis assuming the bimodal mass distribution and the H4 \ac{EoS}.
The next generation instruments \ac{CE} and \ac{ET} detect \ac{BNS} signals at cosmological distances, and nonlocal sources from $z\sim1$ dominate the accumulated information due to volume and redshifting effects.
For detectors that cannot detect the myriad of distant binaries, though, the available \ac{SNR} is dominated by rare loud signals.

The area under the square of each curve in Fig.~\ref{fig:SNRpre_manyIFO} corresponds to the total information, or total \ac{SNR}$^2$, collected during one year of observation.
By comparing the relative information gained between different detectors we can gauge their respective contributions towards a characterization of the \ac{NS} \ac{EoS}. 
Table~\ref{tab:compIFO} presents the relative information gained from each detector compared to \ac{CE} $40\mathrm{km}$ for pre-merger ($D=\{1,2\}$) and post-merger signals.
The ratios amount to years of observation with each detector to match one year of \ac{CE} $40\mathrm{km}$ observation, or equivalently how many detectors of a certain class would be required to match a single \ac{CE} $40\mathrm{km}$ instrument.

Third generation detectors offer at least an order of magnitude more information about both pre-merger and post-merger signals compared to non-cosmological \ac{BNS} detectors.
The increased high-frequency sensitivity for the $20\mathrm{km}$ \ac{CE} configuration results in approximately similar performance compared to its $40\mathrm{km}$ counterpart for post-merger signals, though not for pre-merger ones. ``2.5"G detectors such as OzHF and Voyager offer a clear improvement over aLIGO, and could potentially lead to a post-merger detection (total SNR $\gtrsim5$) and stringent pre-merger EoS constraints (total SNR $\gtrsim 33\,(60)$ with $D=1$ and $\gtrsim 16\,(15)$ with $D=2$ for OzHF (Voyager)).  

\begin{table}
\centering
\begin{tabular}{|c|c|c|c|} 
\hline
Detector & \begin{tabular}[c]{@{}c@{}}Pre-merger, \\~D=1 \end{tabular} & \begin{tabular}[c]{@{}c@{}}Pre-merger, \\~D=2 \end{tabular} & \begin{tabular}[c]{@{}c@{}}Post-merger, \\~D=1\end{tabular}  \\ 
\hline
CE 40km  & 1  & 1 & 1    \\
ET   & 3   & 2 & 9   \\
CE 20km  & 10 & 12  & 3   \\
Voyager  & 830  & 1400   & 1200   \\
OzHF  & 2700 & 1100   & 450   \\
aLIGO & 64000    & 170000  & 42000  \\
\hline
\end{tabular}
\caption{
Relative information $\sim \mathrm{SNR}^{2D}$  gained from each detector compared to \ac{CE} $40\mathrm{km}$. 
We use results with the bimodal mass distribution and \acp{EoS} H4 and EoS7 for the pre-merger and post-merger signals respectively. 
}
\label{tab:compIFO}
\end{table}

\section{Discussion} 

Over the next decade the sensitivity of \ac{GW} detectors will increase, allowing for detection of a large population of \acp{BNS} at cosmological distances.
These observations are aided by the redshift dependent merger rate as well as cosmological redshifting of the observed signals.
With a conservative detection threshold of \ac{SNR}$>8$ in one $40\mathrm{km}$ CE detector, we find $\sim1/3$ of all \acp{BNS} in the Universe are individually detectable. 
In terms of \ac{SNR}, they correspond to about $60\%$ of the total available \ac{SNR} from all mergers, as the  undetected signals are intrinsically weak and subdominant thanks to the $N^{1/2}$ \ac{SNR} scaling.
The detection procedure deployed for real \ac{GW} observations is more sophisticated~\cite{Cannon:2011vi,Meacher:2015rex}, including a global network of \ac{GW} detectors and potentially being able to characterize not only the individually resolvable signals but also a stochastic background~\cite{Abbott:2017xzg,Vivanco:2019lcp,Sachdev:2020bkk}.
These effects could increase the available information.

Measurement of a single scale of the \ac{NS} \ac{EoS} (such as the neutron star radius or the frequency of post-merger oscillations) is dominated by information from the population of \acp{BNS} at $z\sim1$ for third-generation \ac{GW} detectors.
Current facilities, on the other hand, are inherently insensitive to such a distant source population and constraints are dominated by nearby sources.
This suggests that EoS constraints benefit from both high and low frequency sensitivity: the high part offers direct information about 
finite-size effects, and the low part contributes towards the detection of the distant BNS population in the first place.
Extraction of more than one \ac{EoS} parameter, for example to detect the presence of a phase transition~\cite{Han:2018mtj,Bauswein:2018bma}, 
is also primarily achieved by the few \ac{BNS} sources at high individual \ac{SNR} in the local Universe.
The cosmological reach of third-generation detectors and the large recovered \ac{SNR} for the tidally-corrected signal are essential for simultaneous \ac{EoS} and cosmological inference even beyond the Hubble parameter~\cite{Messenger:2011gi,Messenger:2013fya,DelPozzo:2015bna}. 

The large total \ac{SNR} available to next generation detectors raises the bar for mitigating systematic biases.
Whether the \ac{SNR} comes from a single loud source~\cite{Dudi:2018jzn,Samajdar:2018dcx,Purrer:2019jcp} or is the cumulative result of many weak sources~\cite{Purrer:2019jcp}, accurate waveform models describing any individual \ac{BNS} signal are essential
for unbiased estimates of the system parameters.
The post-merger signal is amenable to morphology independent studies with less stringent accuracy requirements~\cite{Chatziioannou:2017ixj}.
In parallel, interpreting tidal and post-merger measurements hinges on \ac{EoS} representations or universal relations. 
Ongoing work on improved waveform models, numerical simulations, and understanding of the \ac{EoS} is essential for achieving the full potential of next-generation \ac{GW} detectors.

\begin{acknowledgments}
The authors would like to thank Tom Callister for useful discussions.
We also thank Daniel Brown for providing the OzHF sensitivity curve, and Evan Hall for providing the CE 20km sensitivity curve.
C.-J.H. acknowledge support of the National Science Foundation, and the LIGO Laboratory. 
A.B. acknowledges support by the European Research Council (ERC) under the European Union's Horizon 2020 research and innovation programme under grant agreement No. 759253. A.B. and by Deutsche Forschungsgemeinschaft (DFG, German Research Foundation) -- Project-ID 279384907 -- SFB 1245 and Project-ID 138713538 -- SFB 881 (``The Milky Way System'', subproject A10).
J.A.C. acknowledge support of the National Science Foundation Grants PHY-1700765, OAC-1841530 and PHY-1809572 .
LIGO was constructed by the California Institute of Technology and Massachusetts Institute of Technology with funding from the National Science Foundation and operates under cooperative agreement PHY-1764464.
The Flatiron Institute is supported by the Simons Foundation. 
The authors are grateful for computational resources provided by the LIGO Laboratory and supported by National Science Foundation Grants PHY-0757058 and PHY-0823459.
This analysis was made possible by the {\tt LALSuite}~\cite{lalsuite}, {\tt astropy}~\cite{Robitaille:2013mpa, Price-Whelan:2018hus}, {\tt numpy}~\cite{numpy}, {\tt SciPy}~\cite{Virtanen:2019joe} and {\tt matplotlib}~\cite{Hunter:2007ouj} software packages. 
This article carries LIGO Document Number LIGO-P2000143.
\end{acknowledgments}

\bibliography{refs_smaller.bib,refs.bib}
\end{document}